\documentclass[aps,prd,showpacs,scriptaddress,twocolumn,nofootinbib]{revtex4}

\usepackage{amssymb,amsmath,amsfonts}
\usepackage[usenames]{color}
\usepackage[pdftex]{graphicx}
\usepackage[english]{babel}
\usepackage{amsmath,amssymb}
\usepackage{multirow}
\usepackage{txfonts}
\usepackage[percent]{overpic}
\usepackage{overpic}

\newcommand{\beq}{\begin{equation}}
\newcommand{\eeq}{\end{equation}}
\newcommand{\bea}{\begin{eqnarray}}
\newcommand{\eea}{\end{eqnarray}}

\newcommand{\ea}{\end{align}}

\newcommand{\bma}{\begin{pmatrix}}
\newcommand{\ema}{\end{pmatrix}}

\newcommand{\WVU}{\affiliation{Department of Physics, West Virginia University, PO Box 6315, Morgantown,
West Virginia 26506, USA}}
\newcommand{\BNU}{\affiliation{Gravitational Wave and Cosmology Laboratory, Department of Astronomy, Beijing Normal University, Beijing 100875, China}}

\begin{document}

\title{Accumulative coupling between magnetized tenuous plasma and gravitational waves}

\begin{abstract}
We explicitly compute the plasma wave (PW) induced by a plane gravitational wave (GW) travelling through a region of strongly magnetized plasma, governed by force-free electrodynamics. The PW co-moves with the GW and absorbs its energy to grow over time, creating an essentially force-free counterpart to the inverse-Gertsenshtein effect. The time-averaged Poynting flux of the induced PW is comparable to the vacuum case, but the associated current may offer a more sensitive alternative to photodetection when designing experiments for detecting/constraining high frequency gravitational waves. 
Aside from the exact solutions, we also offer an analysis of the general properties of the GW to PW conversion process, which should find use when evaluating electromagnetic counterparts to astrophysical gravitational waves, that are generated directly by the latter as a second order phenomenon. 
\end{abstract}

\author{Fan Zhang} \BNU \WVU

\maketitle
\section{Introduction}
The Laser Interferometry Gravitational-wave Observatory's successful detection \cite{2016PhRvL.116f1102A} of gravitational waves (GW) in the tens to thousands Hertz frequency range heralds in the era of gravitational wave astronomy, allowing us to probe deeper into the depth of the cosmos and the core regions of violent astrophysical events. Just as with electromagnetic observations, GW messengers populate a broad frequency range, and projects are currently underway to detect them in the segments of: around $10^{-16}$Hz through the B-mode polarization of the Cosmic Microwave Background, nanohertz band via pulsar timing arrays, millihertz range by space-based laser interferometers, and upto about $10^4$ Hertz with a network of second generation ground-based interferometers. 

In comparison, activities in the higher frequency ($>100k$Hz) end of the spectrum have been relatively subdued. One of the reasons is that the relevant GW sources have been less certain, so there is a chance that this corner of the GW universe is simply quiet. However, most of the speculative sources proposed so far are intimately tied into fundamental physics (e.g. of cosmological \cite{Grischuk,Maggiore00,1999PhRvD..60l3511G,2007PhRvL..98f1302G,Easther07,Caprini09a,Copeland09,
Caldwell96,Leblond09} and braneworld \cite{Seahra05,Clarkson07} origins), and so a detection in this frequency regime may yield important insights, while null results would still be interesting in terms of constraining exotic models.  Another reason for the lack of interest is that our technology has not been sufficiently advanced to make the current detector designs sensitive enough to make the detection of such speculative sources likely (see e.g. Ref.~\cite{2008PhRvL.101j1101A,2006CQGra..23.6185C}). However, rapid progresses in potentially relevant experimental capabilities are being made in the fields of controlled fusion and laboratory astrophysics. In addition, the wavelengths of high frequency gravitational waves (HFGW) are such that they impose less of a demand on the physical size of the detectors. Therefore, the cost-benefit ratio may eventually justify a new generation of detectors being built to listen for such signals, and it is consequently useful to maintain an active investigation into their design options (see e.g.~\cite{TiltInPrep}). 

As high HFGWs need to be converted into e.g., electromagnetic (EM) signals, in order for us to take a readout, the very core of the design problem is then to find a physical process that achieves this conversion most efficiently. The options explored so far concentrate on GW interacting with a static magnetic field, with or without a background electromagnetic wave (EMW) at the same frequency as the GW \cite{Gertsenshtein62,1966PhRv..142..825Z,1972GReGr...3..401B,1983MNRAS.204..485C,2000CQGra..17.2525C,2006CQGra..23.6185C,Li:2009zzy}. 
In such investigations, the presence of the magnetic field is vital for mediating the coupling between the GW and the EMW, and a curved background spacetime can also serve as the catalyst. More interestingly for our present study though, it has been noted that currents enabled by the presence of a plasma can also greatly enhance the coupling \cite{1983PhRvD..28.2382M,1990ApJ...362..584M,1998CQGra..15.3655G,1999PhRvL..82.3012B,2000ApJ...536..875M,2000PhRvE..62.8493S,2001A&A...377..701P,2003PhRvD..68d4017S,Duez2005}, through their being disturbed by the GW. It is therefore interesting to examine the possibility of replacing the vacuum electromagnetic field with a strongly magnetized tenuous plasma as the quiescent configuration (in a stationary solution of the so-called force-free electrodynamics, thus no background radiation). In previous studies involving magnetized plasma/fluid, while detailed equations of motion are written down and some estimates on the amplitude of induced plasma waves (PW) are made, analytical solutions relevant for HFGW detection are generally lacking, and a precise evaluation of the coupling strength and a depiction of the characteristics, such as the associated charges and currents, of the induced radiation remain illusive (we note that explicit solutions are given by Ref.~\cite{2003PhRvD..68d4017S} for the EMW to GW conversion, but not vice versa, which is noted to be more complicated, and Ref.~\cite{Duez2005} offers standing but not travelling wave solutions). 

In this paper, we leverage some recent advances in modelling plasma dynamics in curved spacetimes to find simple and explicit analytical solutions, and show that unfortunately, the temporally averaged Poynting flux associated with the PW induced by the HFGW is no stronger than their vacuum counterparts. The introduction of the plasma however allows for the existence of currents, whose amplitudes depend on that of the GW linearly, and makes possible more sensitive detectors of the ammeter type. It is beyond the scope of this paper to design a technologically viable detector though, and therefore the parameter choices are somewhat arbitrary. We hope our demonstration of the potential enhancement in sensitivity through plasma injection, as well as the introduction of the user-friendly analytical solutions, would solicit interest from experts and lead to more detailed studies. We also mention that although not discussed in any detail below, our analytical solutions may also be applied to the beginning of the GW's journey, namely to study their EM counterparts generated by the GWs themselves inside magnetospheres of compact objects. A large body of excellent literature (see last paragraph for references) already exist on this subject, and we refer interested readers to them for potential applications of our results. 

We derive the equations governing the coupling between the GW and the PW in Sec.~\ref{sec:Eqs}, and analyse properties of the conversion process that they encode in Sec.~\ref{sec:Ana}. In particular, we clarify what kinds of GW excite which types of PW, providing discussions from a physical point of view. We then discuss general solutions to these equations in Sec.~\ref{sec:Gen}, and concentrate on a particular one that's relevant for HFGW detection in Sec.~\ref{sec:Sol}. We further sketch the case for a potential detector design in Sec.~\ref{sec:Det}, before concluding with a discussion in Sec.~\ref{sec:Dis}. Unless otherwise stated, the formulae below are expressed in geometrized units where $G=c=\epsilon_0=1$. Boldface letters are used to represent three or four dimensional vectors and tensors, with specific assignments made clear from context. 

\section{The force-free equations} \label{sec:Eqs}
Analytical solutions describing the interaction between a GW and a magnetized fluid has been worked out in Ref.~\cite{Duez2005} for the case of a standing GW, and used to test a magnetohydrodynamics code. For the sake of HFGW detection, we need a description of induced PW converted from a travelling GW, and we solve for it under the assumption that the plasma's mass density is negligible in its contribution to the overall stress-energy tensor as compared to the electromagnetic field itself, or in other words we assume that the plasma is tenuous. Such a situation is described by the so-called ``force-free electrodynamics", frequently invoked when examining astrophysical environments \cite{Goldreich:1969sb,1977MNRAS.179..433B}. 
More specifically, the inertia-less plasma particles do not have a tendency to preserve their previous states of motion, and thus require no separate kinetic equations (the state of the magnetized plasma is sufficiently described with only the electric and magnetic fields). A plasma particle's movement is instead governed entirely by the requirement that it experiences vanishing 4-force density (or else it would be infinitely accelerated), i.e.  
\bea \label{eq:FFECond}
F_{ab}j^b =0 \,,
\eea
where $j^b$ is the 4-current density due to the plasma particles' motion. Enforcing this condition on the currents that act as the source terms in the usual Maxwell's equations leads to the force-free equations, in which the presence of the plasma manifests as nonlinear modifications \cite{Zhang:2015aga}. 

In this particular limit of magnetohydrodynamics, the propagation speed of the plasma waves are that of the speed of light, therefore optimal for creating resonant conditions where phase coherence with the GW is maintained for as long as possibly, allowing for consistent draining (as opposed to periodically feeding back in) of energy from GW to generate as large an EM signal as possible. Adopting the force-free assumption also allows us to take advantage of some technologies that have recently become available \cite{Gralla:2014yja}, resulting in our being able to find closed-form solutions. 

Specifically, let us assume that the spacetime is initially flat and there is an uniform magnetic field along the $z$ direction, which is described by the field 2-form (the Faraday tensor)
\beq \label{eq:BG}
F_0 = B_0 dx \wedge dy\,.
\eeq
In the flat spacetime, the background solution as given by Eq.~\eqref{eq:BG} carries no current, so the force-free condition is satisfied trivially. However, once metric perturbations are introduced (metric becomes $\eta_{ab}+h_{ab}$, where $\eta_{ab}$ denotes the Minkowski values), a background current appears, taking the value of 
\bea \label{eq:BGCurrent}
j_0^a = B_0\bma 
\partial_y h_{tx}-\partial_x h_{ty} \\
\partial_t h_{ty}-\partial_z h_{yz} -\frac{1}{2}\partial_y\left( h_{tt}+h_{xx}+h_{yy}-h_{zz}\right) \\
-\partial_t h_{tx}+\partial_z h_{xz} +\frac{1}{2}\partial_x\left( h_{tt}+h_{xx}+h_{yy}-h_{zz}\right) \\
\partial_x h_{yz}-\partial_y h_{xz}
\ema \,. 
\eea
Consequently, $F_{0ab} j_0^b \neq 0$ in a curved spacetime and that Eq.~\eqref{eq:BG} ceases to be a valid force-free solution. The PWs then emerge to restore force-freeness, and the now PW-added field $2$-form can be written as 
\beq \label{eq:Faraday}
F = B_0 \,d\,\Big(x+ \alpha(t,x,y,z)\Big) \wedge d\,\Big(y+ \beta(t,x,y,z)\Big)\,,
\eeq
whereby the terms $x+\alpha$ and $y+\beta$ are called Euler potentials. That such a decomposition of the field 2-form is possible is established in Refs.~\cite{1997PhRvE..56.2181U,1997PhRvE..56.2198U,Gralla:2014yja}. The force-free equations of motion can then be transcribed into the exterior calculus language as  \cite{Gralla:2014yja}
\bea \label{eq:FFEEqRaw}
(dx+ d \alpha )\wedge d*F=0, \quad 
(dy+ d \beta)\wedge d*F=0\,,
\eea
wherein the spacetime curvature enters only through the Hodge dual operator $*$ (see Appendix \ref{sec:AppHodge} for details), a fact that significantly simplifies formalism.
To make further progress, let us define, as in \cite{2016ApJ...817..183Y}, the auxiliary variables
\bea \label{eq:Defpsi}
\psi_1 = \partial_x \beta -\partial_y \alpha, \quad \psi_2 = \partial_y \beta+ \partial_x \alpha, 
\eea
turning the explicit form of the equations \eqref{eq:FFEEqRaw} (keeping to linear order in metric perturbation) into 
\begin{align}\label{Eq:psiEqs}
 \left(-\partial^2_t+\partial^2_z\right)\psi_1 =& \frac{\partial^2 h_{tx}}{\partial t \partial y}-\frac{\partial^2 h_{ty}}{\partial t \partial x}+\frac{\partial^2 h_{yz}}{\partial z \partial x}-\frac{\partial^2 h_{xz}}{\partial z \partial y}\,, \nonumber \\
\left(-\partial^2_t+\partial^2_x+\partial^2_y+\partial^2_z\right) \psi_2 =&
\frac{1}{2} \left(\partial^2_x+\partial^2_y\right)\left(h_{tt}+h_{xx}+h_{yy}-h_{zz}\right)
\notag\\
& 
+\frac{\partial^2 h_{yz}}{\partial y \partial z}
+\frac{\partial^2 h_{xz}}{\partial x \partial z}-\frac{\partial^2 h_{yt}}{\partial y \partial t}-\frac{\partial^2 h_{xt}}{\partial x \partial t}\,.
\end{align}
The quantity $\psi_1$ that propagates along the $z$ direction (see the left hand side of Eq.~\ref{Eq:psiEqs}) then depicts waves climbing the magnetic field lines or the Alfv\'en waves, while $\psi_2$ describes the fast-magnetosonic waves \cite{2016ApJ...817..183Y}. 

For the purpose of our study, we specialize to a sinusoidal plane gravitational wave, with a uniform transverse profile, propagating in a direction in the $x-z$ plane that extends an angle $\chi$ with the $z$ axis (for a more generic wave profile, both along and transverse to the propagation direction, see Appendix \ref{sec:GenericTrain}). 
We define the amplitudes $h_{\times}$ and $h_+$ for the cross and plus polarizations (and will use $h$ when distinguishing between them is not necessary), such that in an adapted coordinates system $(t, x',y,z')$ where the GW travels along the $z'$ axis (i.e. spatially rotated against the original coordinates around the $y$ axis by the angle $\chi$), the metric perturbation takes on the familiar form of
\bea
h_{a'b'} =\bma 
 0 & 0 & 0 & 0 \\
 0 & h_+ & h_{\times} & 0 \\
 0 & h_{\times} & -h_+ & 0 \\
 0 & 0 & 0 & 0 
\ema
\cos\Big(\phi_0-\omega(t-z')\Big)\,.
\eea
Transferring back to the original coordinate system where the magnetic field is along the $z$ axis, and in which we will carry out our computations, we then have 
\bea \label{eq:GWperb}
h_{ab} = \bma
 0 & 0 & 0 & 0 \\
 0 & h_+\cos ^2\chi   & h_{\times} \cos \chi  & (h_+/2)\sin 2\chi  \\
 0 & h_{\times} \cos \chi   & -h_+ & h_{\times} \sin \chi   \\
 0 & (h_+/2)\sin 2\chi  & h_{\times} \sin \chi   & h_+ \sin ^2\chi  \ema \cos\xi\,,
\eea
where $\xi \equiv \phi_0-\omega(t +x \sin \chi- z \cos \chi )$. The force free equations \eqref{Eq:psiEqs} then become
\begin{align}
&-\frac{\partial ^2\psi _1}{\partial t^2}+\frac{\partial ^2\psi _1}{\partial z^2} \label{eq:FFE1}
=
 h_{\times} \omega ^2\sin ^2 \chi   \cos \chi   \cos \xi\,, \\
&-\frac{\partial ^2\psi _2}{\partial t^2}+\frac{\partial ^2\psi _2}{\partial x^2}+\frac{\partial ^2\psi _2}{\partial y^2}+\frac{\partial ^2\psi _2}{\partial z^2}
=
h_{+} \omega ^2\sin ^2 \chi   \cos  \xi\,. \label{eq:FFE2}
\end{align}
Note that in our derivations above, we have ignored the back-reaction of the electromagnetic field on the metric, which is suppressed by a factor of $G/c^4$ in SI units (and by a corresponding suppression of the EM field strengths when transferring into geometrized units). 

\section{The selection rules} \label{sec:Ana}
We see that the Euler potential formalism allows us to write down very simple equations \eqref{eq:FFE1} and \eqref{eq:FFE2}, from which we can glean answers to important questions such as which types of GW excite which types of PW. We in fact have a fairly clean dichotomy: 
cross-polarized GWs excite the Alfv\'en waves, while 
plus-polarized GWs excite fast-magnetosonic waves. 

The intuitive reasons behind these simple rules are encoded in Eq.~\eqref{eq:BGCurrent}, which represents the EM effects of immersing the background magnetic field in a curved spacetime. Such effects are the intermediate agents responsible for driving the PWs. 
More precisely, the force-free condition demands that
(to leading order in metric perturbation, and thus also $\alpha$ and $\beta$) 
\bea \label{eq:CurrentCancellation}
F_{0ab} j^{(1)b} = F_{0ab} \left(\delta j^b + j_0^b\right) = 0\,,
\eea
where $\delta j =* (d*\delta F)$, with 
\bea \label{eq:deltaF}
\delta F \equiv B_0(dx\wedge d\beta+ d\alpha \wedge dy)\,
\eea
being the leading order perturbation to the field two-form. In essence then, PWs need to be produced in order to provide a current $\delta j^a$ that cancels out (the troublesome components of) the GW-generated background $j_0^a$. 
Furthermore, we note that $j_0^{z}$ is not an active component in Eq.~\eqref{eq:CurrentCancellation}, as $F_{0xy}=-F_{0yx}$ are the only non-vanishing components of ${\bf F}_0$, so only $j_0^{x}$ and $j_0^{y}$ need to be neutralized (currents in the $x-y$ plane experience a Lorentz force from the background magnetic field in the $z$ direction). In contrast, the component $j^{(1)z}$ does not need to vanish, even at leading order, a fact that we will use later to propose HFGW detectors of the ammeter type. The fast-magnetosonic and Alfv\'en waves split the task of current-neutralization between them. Comparing Eq.~\eqref{eq:BGCurrent}
with Eq.~\eqref{Eq:psiEqs}, we see that the source to the fast-magnetosonic wave $\psi_2$ is simply $(\partial_x j_0^y - \partial_y j_0^x)/B_0$, while the source to the Alfv\'en wave $\psi_1$ is $-(\partial_x j_0^x + \partial_y j_0^y)/B_0$. We examine the two types of waves in turn.

The general characteristic of a magnetosonic wave in any magnetized plasma is that it has a tendency (but not absolutely enforced by the operator on the left hand side of Eq.~\eqref{eq:FFE2}) to travel in the direction perpendicular to the background magnetic field ${\bf B}_0$, with its associated perturbation to the magnetic field $\delta {\bf B}$ having a component along ${\bf B}_0$ (see e.g.~\cite{PlasmaBook} and also Eq.~\eqref{eq:MagB} below).
Therefore, $\delta B^z$ is a good quantitative representation for the fast-magnetosonic waves. Substituting into Eq.~\eqref{eq:deltaF} the two terms $\partial_y\beta$ and $\partial_x \alpha$ of Eq.~\eqref{eq:Defpsi} that combine into $\psi_2$, we see that they in fact give us $\delta B^z = B_0 \psi_2$ (obeyed by our specific fast-magnetosonic wave solution presented in Sec.~\ref{sec:Sol}, see Eqs.~\eqref{eq:Solpsi2} and \eqref{eq:MagB} below). 

With regard to polarization, because $\psi_2$ is essentially $\delta B^z$, the GWs effective at inducing fast-magnetosonic waves would be the ones that are capable of coupling to the background magnetic field in such a way as to generate a $\delta B^z$ by introducing currents onto the $x-y$ plane. From Eq.~\eqref{eq:BGCurrent}, the diagonal entries in $h_{ab}$ are obviously adapt at this task, and according to Eq.~\eqref{eq:GWperb}, they correspond to the plus polarization. There are also other terms in $j_0^x$ and $j_0^y$, including a $h_{yz}$ that corresponds to the cross polarization. However, with our long wave train without $y$ dependence, the $j_0^x$ introduced by this term has no variation in the $y$ direction and is thus incapable of producing $\delta B^z$. When a more complicated wave profile is introduced, such as that in Eq.~\eqref{eq:GenericFFE2} of Appendix \ref{sec:GenericTrain}, both polarizations are activated, but the general rule still applies, namely that 
\begin{itemize}
\item GWs effective at introducing a current that rescales the background magnetic field would be efficient in inducing fast-magnetosonic waves. 
\end{itemize}

The Alfv\'en waves on the other hand, are restricted to propagate along the background magnetic field (see the left hand side of Eq.~\eqref{eq:FFE1}), and are characterized by the presence of an accompanying dynamical current component flowing in the same direction (see e.g.~\cite{2016arXiv160309693G,Brennan:2013jla,Yang:2014zva}). 
Therefore, GWs that can produce a dynamical current (as opposed to a constant flux) in the $z$ direction are better ``impedance matched'' 
and more effective at feeding energy into Alfv\'en waves.
Indeed, although the source term to $\psi_1$ in Eq.~\eqref{Eq:psiEqs} is derived by combining $j_0^x$ and $j_0^y$, the particularities of the combination is such that the source term turns out to be simply $\partial_z j^z_0$ (ignoring the shift coordinate freedom terms like $h_{tx}$ that do not appear for GWs, see Eq.~\eqref{eq:GWperb}). From Eqs.~\eqref{eq:BGCurrent} and \eqref{eq:GWperb}, we see that $h_{\times}$ entering through $h_{yz}$ satisfies this requirement, with the lack of $y$ dependence once again suppressing the other polarization hidden in $h_{xz}$. Just as with the fast-magnetosonic waves, more complicated wave profiles allowed in Eq.~\eqref{eq:GenericFFE1} re-activates $h_{+}$. In summary, a more general rule is that 
\begin{itemize}
\item GWs that introduce variable currents along the background magnetic field direction would be more efficient at eliciting Alfv\'en waves.
\end{itemize}

\section{The general solutions} \label{sec:Gen}
We hope that the intuitive guidelines of Sec.~\ref{sec:Ana} would prove useful in more complicated situations, e.g., ones involving secondary EM radiation coming from compact celestial objects. In such cases, the $j_0^a$ currents from immersing a background EM field in curved spacetimes are straightforward to compute as well, without the need to solve any equations (they are simply $*(d*F_0)$). One can then apply the rules of Sec.~\ref{sec:Ana} to predict the GW-induced PW content without having to solve the FFE equations. Nevertheless, for more quantitative predictions, detailed solutions are sought, and we present a recipe for acquiring general solutions in this section. 

We begin by noting that the selection rules do not mean a particular type of modes can not exist in the absence of the correct type of GW. In the absence of a source term, the homogeneous force-free equations can still be solved, and the resulting $\psi_1$ or $\psi_2$ represent waves being injected at the boundaries of the magnetized regions, and simply traverse such regions as their flat-spacetime counterparts would, without drawing energy from the GW. Such solutions are useful when superposed onto the so-called particular solutions to the inhomogeneous equations when constructing general solutions. The inhomogeneous solutions are more complicated to obtain, but fortunately, we have treated the relativistic effects perturbatively. Specifically, the metric perturbations have all been placed onto the right hand side of the force-free equations, leaving us with simple flat spacetime partial differential operators for the principal part of these equations, for whom the Green's functions are well-known. This allows us to adopt standard Green's function methods to solve them, not only for the long sinusoidal wave trains as assumed for Eqs.~\eqref{eq:FFE1} and \eqref{eq:FFE2}, but also the more complicated cases of Appendix \ref{sec:GenericTrain}.  

Let the source terms on the right hand sides of Eqs.~\eqref{eq:FFE1} and \eqref{eq:FFE2} be denoted by $\mathcal{S}_1$ and $\mathcal{S}_2$, then the particular solutions to the inhomogeneous equations are 
\bea \label{eq:Green}
\psi_1({\bf x'}) &=& \int \mathcal{G}_1(\Delta'_t,\Delta'_z) \mathcal{S}_1({\bf x''})\ dt''\, dz'' \,,\\
\psi_2({\bf x'}) &=& \int  \mathcal{G}_2({\bf x''}-{\bf x'}) \mathcal{S}_2 ({\bf x''}) d^4 x''\,, 
\eea
where $\Delta'_z \equiv z'-z''$, $\Delta'_t \equiv t'-t''$, and $\mathcal{G}_{1/2}$ are the Green's functions. The integrations are to be carried out over the entire vessel containing the magnetized plasma. Homogeneous solutions (satisfying the FFE equations with $\mathcal{S}_{1/2}=0$) can then be superposed onto the results in order to satisfy desired boundary and initial conditions. Because we have decoupled the equations for the Alfv\'en and fast-magnetosonic waves, their boundary conditions can be imposed separately. As already mentioned, these homogeneous solutions are simply freely propagating waves that behave as if they are in flat spacetime, and do not interact with the GW. Therefore, fixing boundary conditions is not important when studying the GW to PW conversion process, and they do not determine which type of PW is induced by the GW. They are however required if one is to compute the observables such as energy fluxes or currents, to be measured by HFGW detectors, because freely propagating waves injected at the boundaries contaminate these quantities. We will discuss this further in Sec.~\ref{sec:Sol}. 
Returning to the particular solution, the Green's functions are those of the flat spacetime, specifically 
\bea
\mathcal{G}_1(z'',t'';z',t')&=&\frac{1}{2}\Theta(\Delta'_t)\Theta(\Delta'_t+\Delta'_z)\Theta(\Delta'_t-\Delta'_z)\,, \label{eq:G1}\\
\mathcal{G}_2({\bf x''}; {\bf x'}) &=& \frac{\delta(t'-(t''-R))}{4 \pi R}\,,
\eea
where $R\equiv |{\bf x}''-{\bf x}'|$, and $\Theta$ are the Heaviside step functions. 

For simpler $\mathcal{S}_{1/2}$ such as those appearing on the right hand sides of Eqs.~\eqref{eq:FFE1} and \eqref{eq:FFE2}, closed-form solutions for $\psi_{1/2}$ exist. We will be examining a closed-form particular solution for $\psi_2$ in Sec.~\ref{sec:Sol} in quite some detail, so here, we present only a solution for $\psi_1$. Assuming the interaction between the GW and the magnetized plasma begins at $t_0$, and that the $z$ extent of the plasma container is sufficiently large that at the time $t'$ concerned, all the regions where $\mathcal{G}_1 >0$ are included within, then we have 
\bea
\psi_1 &=& 2 h_{\times} \Bigg[\sin ^2\frac{\chi }{2} \sin \left(\Delta _0 \omega  \cos ^2\frac{\chi }{2}\right) \sin \left(\hat{\xi}+\frac{1}{2} \omega  \Delta_0 \cos \chi \right)
\notag \\
&&-\cos ^2\frac{\chi }{2} \sin \left(\Delta _0 \omega  \sin ^2\frac{\chi }{2}\right) \sin \left(\hat{\xi}-\frac{1}{2} \Delta _0 \omega  \cos \chi \right)\Bigg]\,, \notag \\
\hat{\xi} &\equiv& \phi_0-\frac{1}{2} \omega  (t_0+t')-x' \omega  \sin \chi +\omega  z' \cos \chi\,,  \label{eq:psi1Sol}
\eea
where also $\Delta_0\equiv t'-t_0$.

Given the solutions for $\psi_{1/2}$, we will then need to reconstruct the original $\alpha$ and $\beta$ perturbation functions before we can compute the EM fields. From Eq.~\eqref{eq:Defpsi}, we deduce that these quantities satisfy the two-dimensional elliptic equations 
\bea
\left(\frac{\partial^2}{\partial x^2}+\frac{\partial^2}{\partial y^2} \right)\alpha &=& -\frac{\partial \psi_1}{\partial y} + \frac{\partial \psi_2}{\partial x}\,, \label{eq:AlphaFrompsi}
\\
\left(\frac{\partial^2}{\partial x^2}+\frac{\partial^2}{\partial y^2} \right)\beta &=& \frac{\partial \psi_1}{\partial x} + \frac{\partial \psi_2}{\partial y}\,. 
\eea
Once again, we have a Green's function giving us the particular solution
\begin{equation}
\alpha =\int\frac{\ln |\rho-\rho'|}{2 \pi} \mathcal{S}_{\alpha}(\rho')\,d^2 \rho \,, 
\end{equation}
where $\rho$ denotes locations on the $x-y$ plane, $\mathcal{S}_{\alpha}$ represents the right hand side of Eq.~\eqref{eq:AlphaFrompsi}, and a similar solution exists for $\beta$. A further integration by parts allows us to write 
\bea
\alpha({\bf x}) &=& \int dx' dy' \frac{\Delta_x\psi_2(x',y',z)-\Delta_y\psi_1(x',y',z)}{2 \pi [\Delta_x^2+\Delta_y^2]}\,, \\
\beta({\bf x}) &=& \int dx' dy' \frac{\Delta_x\psi_1(x',y',z)+\Delta_y\psi_2(x',y',z)}{2 \pi [\Delta_x^2+\Delta_x^2]}\,,
\eea
whereby $\Delta_x\equiv x-x'$ and $\Delta_y \equiv y-y'$. It turns out that symbolic manipulation software such as \verb!Mathematica! are capable of generating closed-form expressions for $\alpha$ and $\beta$ that correspond to the $\psi_1$ given by Eq.~\eqref{eq:psi1Sol}. The expressions are too long and tedious to be reproduced here, but the integrands are straightforward to input. On the other hand, the closed-form fast-magnetosonic wave we examine in the next section will have simple expressions for $\alpha$, $\beta$, as well as the Faraday tensor.

We have (here and in the next section) thus given closed-form particular solutions to both of the inhomogeneous Eqs.~\eqref{eq:FFE1} and \eqref{eq:FFE2}, which can be combined with homogeneous solutions to generate general solutions satisfying different boundary and initial conditions. For more complicated $\mathcal{S}_{1/2}$ than considered in this paper, numerical integrations can be utilized to yield the various quantities.

\section{The accumulative solutions} \label{sec:Sol}
In this section, we specialize to an exact solution that resembles the vacuum inverse-Gertsenshtein effect most closely, and is therefore the most relevant for HFGW detection. 
Eqs.~\eqref{eq:FFE1} and \eqref{eq:FFE2} are simple enough that we can isolate such interesting solutions straight away. We expect efficient GW to PW conversion to be more likely when the two types of waves can co-move in the same direction and  retain a constant relative phase. For the Alfv\'en waves, the intrinsic propagation direction is along the $z$ axis, but the source term on the right hand side of Eq.~\eqref{eq:FFE1} vanishes when $\chi=0$. There is however no restriction on the propagation direction for the fast-magnetosonic waves, and therefore we concentrate on this variety. 
More specifically, we search for solutions that depict the following scenario: that the GW travelling in the direction determined by $\chi$ excites a fast-magnetosonic wave moving in the same direction, which continues to siphon energy off of the GW while it propagates. Indeed, the following ansatz 
\bea \label{eq:Solpsi2Raw}
\psi_2 = \frac{\zeta}{4} h_{+} \omega \sin^2\chi \sin  \xi\,,
\eea
solves Eq.~\eqref{eq:FFE2}, where $\zeta = t-x \sin \chi +z \cos \chi$ is the ``advanced'' time (with $\xi$ being the retarded time) that measures distance along the propagation direction. Therefore, the PW grows linearly in amplitude as it propagates, and it is in this sense we term the solution ``accumulative". 

\begin{figure}[t,b]
\includegraphics[width=0.95\columnwidth]{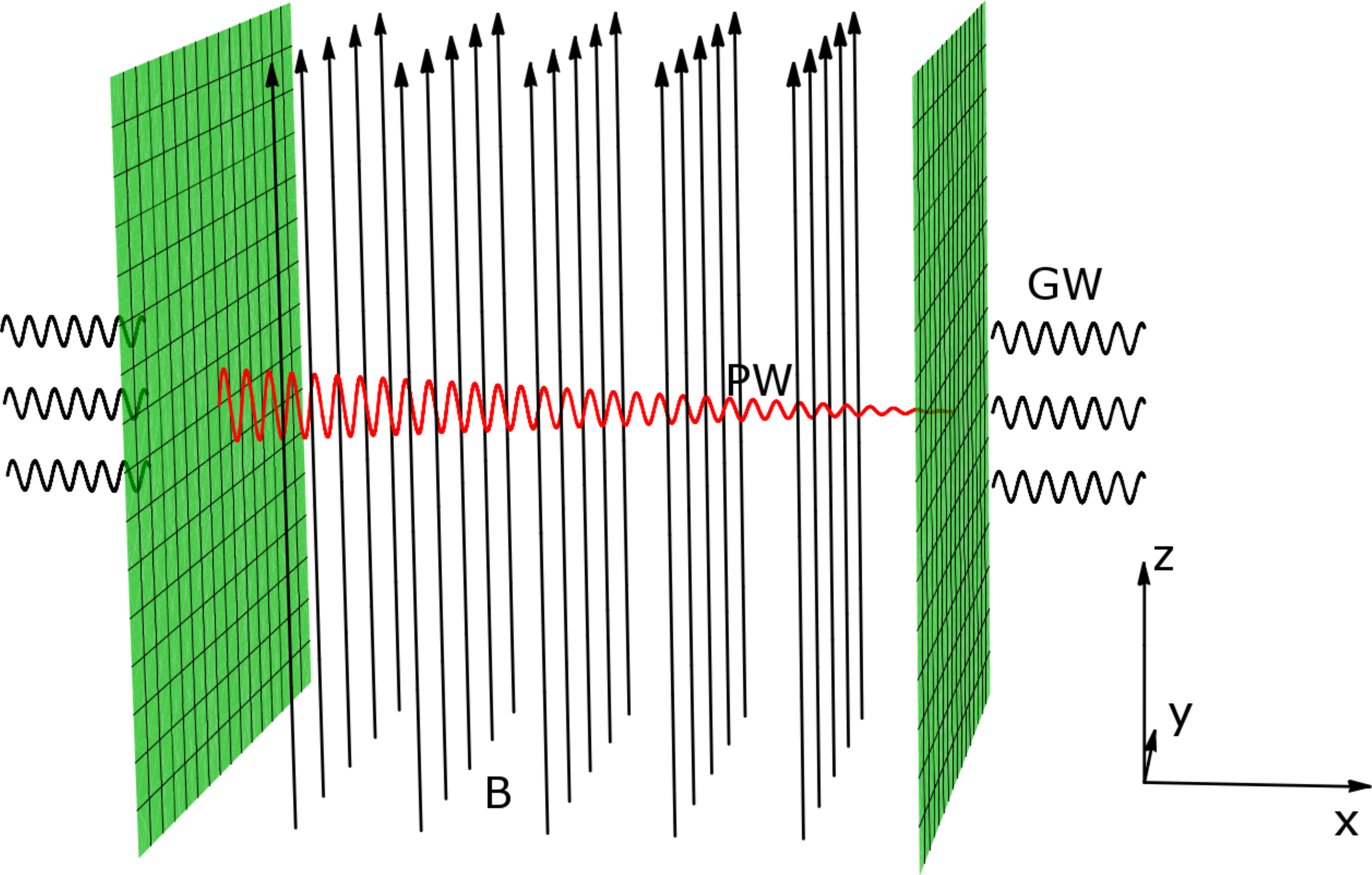}
\caption{A schematic depiction of the interactions. The plasma and a strong magnetic field (black arrows) in the $z$ direction is contained between the two green screens. The GW (black wiggles) traverses the magnetic field orthogonally from the right. A fast-magnetosonic wave (red wiggles) is induced which grows linearly over distance along $-x$, resulting in a small Poynting flux being registered on the left screen.  
}
\label{fig:Setup}
\end{figure}

Below, we will concentrate on this type of growing solutions and elucidate their properties, which would be useful if actual experimental apparatus is to be designed to exploit it in the detection of HFGW. We also note that the source term on the right hand side of Eq.~\eqref{eq:FFE2} is the largest when $\chi=\pi/2$ (GW propagating along the magnetic field generates no current \cite{1983PhRvD..28.2382M,1990ApJ...362..584M}), and we will assume this value for the experimental setup. We will however retain $\chi$ in our formulae in order to present as generic a solution as possible, to facilitate application to other occasions. Schematically, we trap a strongly magnetised plasma in-between two screens as depicted in green in Fig.~\ref{fig:Setup} (for simplicity, we will set $x=0$ on the right screen where the GW first comes into contact with the plasma), and examine the PW generated by the GW in that region. 
For comparison, we mention that a growing vacuum EMW would similarly be induced by the GW if the plasma is evacuated from the magnetized region. This phenomenon is commonly termed the inverse-Gertsenshtein effect \cite{Gertsenshtein62,Lupanov67,1977PhRvD..16.2915D}, and ours is essentially a force-free version of it. 

Before we compute the relevant field quantities, a few remarks regarding the growing solutions for $\psi_2$ are in order. First of all, the advanced time $\zeta$ has an arbitrary initial value. Taking the transformation $\zeta \rightarrow \zeta+\zeta_0$ with a constant $\zeta_0$ will simply introduce an additional contribution that solves the homogeneous part of Eq.~\eqref{eq:FFE2}. Physically, the value of $\zeta_0$ is determined by the initial condition at wherever the GW begins to interact with the plasma, and whether there has already been a seed wave present then. Furthermore, the $\zeta$ in Eq.~\eqref{eq:Solpsi2Raw} can be broken into components, and the segmental expressions falling out such as 
\bea
\psi_2 = \frac{t}{2} h_{+} \omega \sin  \xi \sin^2\chi \,,
\eea
also solve Eq.~\eqref{eq:FFE2}. These solutions represent essentially the same physics of amplitude growth during concurrent GW-PW propagation (along null geodesics of constant $\xi$), but with a rescaling and a geodesic-dependent translation of the affine parameter $\zeta$. Which particular form of the growing solutions to use obviously depends on the boundary conditions we wish to impose, and as we will enforce a no-initial Poynting flux condition on the entrance screen at $x=0$, it turns out that the most convenient choice is of the form
\bea 
\psi_2= -\frac{x}{2}  h_+  \omega \sin \xi \sin\chi\,.
\eea

Note that we have the freedom to add freely propagating waves (solutions to the homogeneous part of Eq.~\ref{eq:FFE2}) to the solution to help enforce boundary and initial conditions. In order to avoid being overly restrictive, given that we do not know of the detailed properties of the would-be detectors, and the initial conditions for their interaction with GWs, we look for steady state solutions representing the plasma after interacting with a long GW train for a significant amount of time, and only impose the boundary condition that the screen at $x=0$ does not inject any Poynting flux into the cavity, so that flux registered on the left screen comes from the GW conversion. 
Because we do not enforce boundary conditions on the other surfaces of the plasma cavity, our solution will not be unique, we instead examine a representative solution, given by 
\bea 
\psi_2= -\frac{1}{2} h_+ \Big(\cos \xi +x \omega  \sin \xi  \sin \chi \Big) \,. \label{eq:Solpsi2} 
\eea

With Eq.~\eqref{eq:Solpsi2} and the assumption that $\psi_1=0$, the solutions to Eq.~\eqref{eq:Defpsi} are easy to obtain by inspection, giving us
\bea \label{eq:alphabeta}
\alpha= -\frac{x}{2} h_+ \cos \xi  \,, \quad  \beta=0\,,
\eea 
which indeed satisfy the original equations \eqref{eq:AlphaEq} and \eqref{eq:BetaEq} for these entities. We caution that solution \eqref{eq:alphabeta} is valid only when either $h_{\times} = 0$, or $\chi=\pi/2$, or $\chi=0$. This is because our solution corresponds to $\psi_1=0$, but a vanishing $\psi_1$ is only a solution to Eq.~\eqref{eq:FFE1} when the right hand side of that equation vanishes. For compactness, we will not display the equations for each case separately, but readers should bear in mind this constraint when applying the formulae below. 

Substituting Eq.~\eqref{eq:alphabeta} into Eq.~\eqref{eq:Faraday} for the Faraday tensor, we can subsequently compute the electric field ${\bf E}$ through 
$E^a = F^{ab}\tau_b$, where $\tau_b$ is the time-like one form orthogonal to spatial slices of constant $t$, and magnetic field $B^d = (1/2)\epsilon^{abcd}F_{ab}\tau_c$, as well as the Poynting vector ${\bf P}={\bf E}\times {\bf B}$ (we adopt the convention that $\epsilon^{0123}=-1$, so $\epsilon^{123}=\tau_a\epsilon^{a123}=1$ in the Minkowski limit). We use perturbed metric for the computation, but keep the results upto only linear order in $h$, which are (all contravariant spatial vectors)
\bea
{\bf E}^{(1)}&=& 
\frac{x}{2} B_0   h_+ \omega \sin \xi 
\bma 0 \\ 1 \\ 0\ema
\,, 
\quad 
{\bf B}^{(0)}= 
B_0\bma 0 \\ 0 \\
1\ema \,,
\label{eq:ElecE} \\
\notag \\
{\bf B}^{(1)}&=& -\frac{1}{2} B_0 h_+ \cos \xi
\bma 0 \\ 0 \\ 1\ema 
-\frac{x}{2} B_0 h_+ \omega  \sin \xi
\bma  \cos \chi \\ 0 \\
\sin \chi \ema
\,, \label{eq:MagB} \\
{\bf P}^{(1)}&=& \frac{x}{2} B_0^2 h_+ \omega  \sin \xi \bma 1\\0\\0 \ema 
\,, \label{eq:Power}
\eea
where we note that for our steady state solution, there is a component in ${\bf B}^{(1)}$ that does not grow linearly in $x$. Nevertheless, there is no first order Poynting flux at $x=0$. In more details, the ${\bf P}^{(1)}$ above is produced by crossing the induced electric field with the background magnetic field . This first order flux oscillates between going in the positive and negative $x$ directions, and would not contribute to a temporally integrated signal. A consistent flux (without periodic sign reversals causing nearly cancelling positive and negative accumulations) associated with a propagating wave on the other hand, comes from the second term of
\begin{align}
\mathcal{P}^{(2)} \equiv {\bf E}^{(1)} \times {\bf B}^{(1)} 
=& -\frac{x}{8} B_0^2 h_+^2 \omega  \sin 2\xi \bma 1 \\ 0 \\0 \ema 
\notag \\ 
&- \frac{x^2}{4} B_0^2  h_+^2 \omega^2 \sin ^2\xi
\bma \sin \chi \\ 0 \\ -\cos \chi \ema\,.
\end{align}
We caution that $\mathcal{P}^{(2)}$ does not constitute the entirety of the second order Poynting flux ${\bf P}^{(2)}$, as there is a contribution from the second order ${\bf E}^{(2)}$ crossing with the background ${\bf B}^{(0)}$. The examination of such a term is beyond the scope of our current investigation, but it may nevertheless contain a part that does not average out over time. 

With PWs, charges and currents are allowed, which can be computed by simply noting that half of the Maxwell's equations are given by $d*{\bf F}={\bf J}$, where ${\bf J}$ is the current 3-form, relating to the usual $4$-current density $j^a$ (zeroth component being the charge density $\rho$) via $j^a = (1/3!)\epsilon^{abcd} J_{bcd}$. To first order in metric perturbation, we have 
\bea \label{eq:Current}
\rho^{(1)} = 0\,, \quad 
{\bf j}^{(1)}= B_0  h_{\times} \omega \sin \xi  \sin \chi  \bma  \cos \chi  \\ 0 \\  \sin \chi \ema \,, 
\eea
where the cross polarization $h_{\times}$ is introduced, and $h_+$ removed, when taking the Hodge dual to obtain $*{\bf F}$. When $\chi=\pi/2$ so we can have $h_{\times} \neq 0$, the nonvanishing last row of ${\bf j}^{(1)}$ is simply $j_0^z$ (see Eq.~\eqref{eq:BGCurrent}) that does not need to be removed by the PWs (see discussion in Sec.~\ref{sec:Ana}). We also do not need to be concerned with $h_{\times}$ generating Alfv\'en waves, as the source term in Eq.~\eqref{eq:FFE1} vanishes when $\chi=\pi/2$.
Now that we have the explicit expression for $j^a$, it is easy to verify that Eq.~\eqref{eq:FFECond} is indeed satisfied. 

\section{The high frequency gravitational wave detector} 
\label{sec:Det}

In the absence of a GW, the plasma will be quiescent, with microwave detectors on the left hand screen registering no Poynting flux. When a GW wave train traverses the magnetic field orthogonally, a fast-magnetosonic wave is produced, which grows in amplitude as it propagates. Such a wave is described by Eq.~\eqref{eq:Solpsi2}.  The dominant instantaneous energy flux recorded on the left screen at $x=-L$ is subsequently given by Eq.~\eqref{eq:Power}, which when translated into SI units becomes (unit of $P$ is watts per square meter)
\begin{align}\label{eq:SignalAmpInst}
P_x^{(1)}=2.0\times 10^{-12} B_{10}^2 L_{10} \omega _{\text{GHz}} h_{+-30} \sin \xi \,,
\end{align}
where 
\bea
B_{10}&=&B_0/(10\text{Tesla})\,, \,\, h_{+-30} = h_+/10^{-30}\,, \,\, h_{\times-30} = h_{\times}/10^{-30}\,, \notag \\  L_{10}&=&L/(10 \text{m})\,, \,\, \omega _{\text{GHz}} = \omega/(5\times 10^9 \text{Hz})\,,
\eea
are dimensionless rescaled quantities normalized by typical values (experimentally reasonable and commonly shared across different HFGW source predictions) \cite{Li:2009zzy}. 
The time-averaged flux due to the propagating wave on the other hand, appears at the next order, and evaluates to
\begin{align}\label{eq:SignalAmpAvg}
\langle \mathcal{P}_x^{(2)}\rangle  =-8.3\times 10^{-41} B_{10}^2 L_{10}^2 \omega _{\text{GHz}}^2 h_{+-30}^2 \sin \chi\,.
\end{align}
We notice that, as typical for HFGW detection, higher frequencies can compensate for low strains \cite{1674-1056-22-12-120402}.

The time-averaged flux \eqref{eq:SignalAmpAvg} is at the $\mathcal{O}(h^2)$ order, thus comparable to the vacuum inverse-Gertsenshtein effect \cite{Li:2009zzy}. In addition, we do not expect the remaining part of ${\bf P}^{(2)}$ from ${\bf E}^{(2)}\times {\bf B}^{(0)}$ to provide significant enhancements, as that term would also be proportional to $B_{10}^2 h_{-30}^2$ and would unlikely contain much higher powers of $\omega_{\text{GHz}}$ ($L_{10}$ shares the same power for dimensional reasons). This is because even with new second order source terms being introduced into the right hand sides of Eq.~\eqref{Eq:psiEqs}, the differential operators would still contain only two derivatives in order for the dimensions to match up, therefore we do not have extra derivatives to bring out additional factors of $\omega$. Furthermore, numerical factors appearing in our geometrized unit computations are all of moderate values, often arising from combinatorics tied down to the $3+1$ dimensionality of our spacetime, so there is unlikely significant boosts from large newly emerging coefficients in the second order computations. In contrast, the state of art photodetector sensitivity circa 2012 is $10^{-22}$W \cite{Cruise:2012zz}. Consequently, detecting this temporally averaged flux is not feasible. 

However, we notice that the instantaneous flux \eqref{eq:SignalAmpInst} is at $\mathcal{O}(h)$ and is comfortably measurable from a purely power amplitude point of view, provided that the detector is located inside of the magnetized region. With previous detector designs searching for GW to vacuum EMW conversions, the photodetectors are placed outside of the screens that are presumed transparent to the EMW (see Fig.~2 in Ref.~\cite{Cruise:2012zz}), even though a similar first order effect should be present in that case as well (see e.g. Eq.~1 of Ref.~\cite{1674-1056-22-12-120402}). This may be due to concerns regarding the effect of the magnetic field on the photodetector (although this issue appears manageable \cite{Beznosko:2005sy}), or uncertainty in whether a photodetector is capable of registering such a rapidly sloshing energy flux. After all, Eq.~\eqref{eq:SignalAmpInst} does not represent a steady stream of photons moving towards either the positive or the negative $x$ direction. 

With PWs however, there is an accompanying current, also at $\mathcal{O}(h)$, due to the kinetic motion of the plasma particles, which may be more readily detectable given that no photon to current conversion is required. Specifically, currents flowing in either direction is permitted, and a direct measurement of the first order effect simply as a high frequency alternating current may be possible. We have computed the current and charge densities for the growing solution, given by Eq.~\eqref{eq:Current}, which in SI units takes the value of (with units of amperes per square meter, and we set $\chi=\pi/2$ so $h_{\times}$ does not need to vanish)
\begin{align}
{\bf j}^{(1)} = 1.3\times 10^{-22} B_{10} h_{\times -30} \omega _{\text{GHz}} 
\sin \xi 
\bma 0 \\ 0 \\ 1 \ema\,. 
\end{align}
The dependence of the current density on $x$ through $\xi$ means that the current collector would likely need to be stratified along the $x$ direction, with signals from adjacent stripes (each of half a wavelength, or $3\pi/(50\omega_{\text{GHz}})$ meters) aggregated by subtraction instead of addition, to ensure that positive and negative contributions in the current density do not cancel out. This may also help subtract out some of the background stray noise currents. If we manage to construct a current collecting area of ten meters by ten meters, and perhaps increase the target frequency range (note $\omega_{\text{GHz}}$ is angular frequency) and/or magnetic field strength, we can bring the total current to the atto ($10^{-18}$) ampere regime, for which measurement equipments are already available at the turn of the century \cite{measurementbook}, and the state of the art may be even more sensitive.  

\section{Discussion \label{sec:Dis}}
In this paper, we have studied the force-free version of the inverse-Gertsenshtein effect, and obtained explicit solutions, allowing us to compute the first order currents that were not present in the vacuum case, offering possibly a new avenue for detecting HFGW. Specifically, we have considered a quiescent background configuration, where the magnetic field is constant in space and time. There is no background electric field, Poynting flux, or current density present in the absence of a GW. We then target nascent electrical current or Poynting flux converted from the GW for detection. 

In other words, our consideration has been restricted to detectors of the ``conversion" type \cite{Cruise:2012zz,Gertsenshtein62,Lupanov67,1970NCimB..70..129B,DeLogi:1977qe,Cruise:2012zz}. Our computation shows that temporally averaged Poynting flux for the induced PW is at the $\mathcal{O}(h^2)$ order, which turns out to be the same for the vacuum cases, and thus the introduction of plasma is unlikely to lead to feasible detectors of the traditional photodetection design. 
In previous literature, more complicated designs termed 
the ``geometric" types \cite{1978PhLA...68..165P,
1972GReGr...3..401B,
1976GReGr...7..583B,
1974NCimB..19..105T,
1966PhRv..142..825Z,
1975PThPh..54.1309T,
1993ApJ...418..202F,
1968AnPhy..47..173C,
1971PhRvL..26.1398B,
1975GReGr...6..329B,
1983MNRAS.204..485C,
2000CQGra..17.2525C,
Cruise06,
Li:2009zzy} have also been proposed. With such detectors, effects depending on the GW amplitude at the first order is seen in e.g., the polarization state or frequency, of a strong background EMW. Such more subtle characteristics of a EMW tends to be relatively difficult to measure however, and the demand on the purity of the background EMW is also high, so one faces not only a detection problem, but also a generation one. In contrast, the current associated with the PW also appears at the $\mathcal{O}(h)$ order, but requires fewer intermediate stages before a signal readout can be taken, in addition to needing no nontrivial backgrounds. So introducing plasma may perhaps lead to more sensitive ammeter designs of the conversion type. 

Although beyond the scope of this paper, one may of course also consider geometric type force-free detectors, where background Alfv\'en or fast-magnetosonic wave pulses are launched through the plasma cavity, and emerge altered by the GW. 
The examination of such design choices should be carried out in conjunction with an investigation into technological options, and the general method introduced here should be adaptable to such studies. 

Lastly, we note that although there is no background current or Poynting flux in theory, the motion of the plasma particles may not conform to force-free electrodynamics perfectly in practical situations, or vibrations of the apparatus may launch unwanted waves (sound wave conversion). Therefore, some level of stray background radiation and current is to be expected, which is likely stochastic. On the other hand, although we have used a clean monochromatic long GW train for analysis, the predicted HFGW signals are also mostly stochastic, and the standard method for analysing such signals is by examining the correlation between readouts from two detectors placed in close proximity \cite{Cruise06}. This poses certain technical challenges, for example, the local vibrations at the two detectors would also be correlated. In addition, the usual gravity gradient noise etc., all need to be carefully investigated, and so in the end, the noise budget, instead of device sensitivity, may well prove to be the limiting factor for the ammeter detectors. 

\acknowledgements
We thank Hao Wen for discussions regarding HFGW detector noises, and an anonymous referee for helpful suggestions that led to the introduction of Secs.~\ref{sec:Ana} and \ref{sec:Gen}. F.~Z.~is supported by the National Natural Science Foundation of China grants 11443008 and 11503003, Fundamental Research Funds for the Central Universities grant No.~2015KJJCB06, and a Returned Overseas Chinese Scholars Foundation grant.

\appendix
\section{The Hodge dual expressions \label{sec:AppHodge}}
When the perturbed metric is $\eta_{ab}+\epsilon h_{ab}$ (we introduce a flag $\epsilon$ to help track the order of small quantities), the Hodge dual rules are
\begin{widetext}
\begin{align*}
&{}^* dt\wedge dx = \frac{1}{2} dy\wedge dz (h_{tt} \epsilon -h_{xx}\epsilon +h_{yy}\epsilon +h_{zz}\epsilon +2)+h_{ty} \epsilon  dt\wedge dz-h_{tz}\epsilon  dt\wedge dy+h_{xy}\epsilon  dx\wedge dz-h_{xz}\epsilon  dx\wedge dy +\mathcal{O}(\epsilon^2),
\\ &
{}^* dt\wedge dy= \frac{1}{2} dx\wedge dz (-h_{tt} \epsilon -h_{xx}\epsilon +h_{yy}\epsilon -h_{zz}\epsilon -2)-h_{tx} \epsilon  dt\wedge dz+h_{tz}\epsilon  dt\wedge dx-h_{xy}\epsilon  dy\wedge dz-h_{yz}\epsilon  dx\wedge dy+\mathcal{O}(\epsilon^2),
\\ &
{}^*dt\wedge dz= \frac{1}{2} dx\wedge dy (h_{tt} \epsilon +h_{xx}\epsilon +h_{yy}\epsilon -h_{zz}\epsilon +2)+h_{tx} \epsilon  dt\wedge dy-h_{ty} \epsilon  dt\wedge dx-h_{xz}\epsilon  dy\wedge dz+h_{yz}\epsilon  dx\wedge dz+\mathcal{O}(\epsilon^2),
\\ &
{}^*dx\wedge dy= \frac{1}{2} dt\wedge dz (h_{tt} \epsilon +h_{xx}\epsilon +h_{yy}\epsilon -h_{zz}\epsilon -2)+h_{tx} \epsilon  dx\wedge dz+h_{ty} \epsilon  dy\wedge dz-h_{xz}\epsilon  dt\wedge dx-h_{yz}\epsilon  dt\wedge dy+\mathcal{O}(\epsilon^2),
\\ &
{}^*dx\wedge dz= \frac{1}{2} dt\wedge dy (-h_{tt} \epsilon -h_{xx}\epsilon +h_{yy}\epsilon -h_{zz}\epsilon +2)-h_{tx} \epsilon  dx\wedge dy+h_{tz}\epsilon  dy\wedge dz+h_{xy}\epsilon  dt\wedge dx+h_{yz}\epsilon  dt\wedge dz+\mathcal{O}(\epsilon^2),
\\ &
{}^*dy\wedge dz= \frac{1}{2} dt\wedge dx (h_{tt} \epsilon -h_{xx}\epsilon +h_{yy}\epsilon +h_{zz}\epsilon -2)-h_{ty} \epsilon  dx\wedge dy-h_{tz}\epsilon  dx\wedge dz-h_{xy}\epsilon  dt\wedge dy-h_{xz}\epsilon  dt\wedge dz+\mathcal{O}(\epsilon^2),
\end{align*}
and the corresponding force-free equations are
\begin{align}
\frac{1}{2}\left(2\frac{\partial ^2\alpha }{\partial x\partial y}+2\frac{\partial h_{\text{ty}}}{\partial t}-\frac{\partial h_{\text{tt}}}{\partial y}-\frac{\partial h_{\text{xx}}}{\partial y}-\frac{\partial h_{\text{yy}}}{\partial y}+\frac{\partial h_{\text{zz}}}{\partial y}-2\frac{\partial h_{\text{yz}}}{\partial z}-2\frac{\partial ^2\beta }{\partial t^2}+2\frac{\partial ^2\beta }{\partial y^2}+2\frac{\partial ^2\beta }{\partial z^2}\right)&=\mathcal{O}(\epsilon^2)\,,
\\ 
\frac{1}{2}\left(2\frac{\partial ^2\beta }{\partial x\partial y}+2\frac{\partial h_{\text{tx}}}{\partial t}-\frac{\partial h_{\text{tt}}}{\partial x}-\frac{\partial h_{\text{xx}}}{\partial x}-\frac{\partial h_{\text{yy}}}{\partial x}+\frac{\partial h_{\text{zz}}}{\partial x}-2\frac{\partial h_{\text{xz}}}{\partial z}-2\frac{\partial ^2\alpha }{\partial t^2}+2\frac{\partial ^2\alpha }{\partial x^2}+2\frac{\partial ^2\alpha }{\partial z^2}\right)&=\mathcal{O}(\epsilon^2)\,.
\end{align}
\end{widetext}
With the definition of $\psi_1$ and $\psi_2$ as in Eq.~\eqref{eq:Defpsi}, these equations simplify to Eq.~\eqref{Eq:psiEqs}. We can also specialize the source terms to the plane GW considered in the main text, in which case the equations satisfied by $\alpha$ and $\beta$ to $\mathcal{O}(h)$ are
\bea
-\frac{\partial ^2\alpha }{\partial x\partial y}+\frac{\partial ^2\beta }{\partial t^2}-\frac{\partial ^2\beta }{\partial y^2}-\frac{\partial ^2\beta }{\partial z^2} &=& \frac{h_{\times}}{2} \omega  \sin \xi  \sin 2\chi   \,, \label{eq:AlphaEq}\\
\frac{\partial ^2\beta }{\partial x\partial y}-\frac{\partial ^2\alpha }{\partial t^2}+\frac{\partial ^2\alpha }{\partial x^2}+\frac{\partial ^2\alpha }{\partial z^2} &=& -h_+\omega  \sin \xi \sin \chi \,. \label{eq:BetaEq}
\eea

\section{Generic wave profiles \label{sec:GenericTrain}}
In the main text, we have specialized to an infinite plane wave train with a sinusoidal profile in the propagation direction, and no variation across the wavefront in the transverse directions. Such configurations are the most natural for HFGW detection, but when it comes to examining the accompanying EM radiation generated by an astrophysical GW, it is desirable to expand our consideration to more generic wave profiles. To this end, we define new variables 
($\xi_3$ is the retarded time, while $\xi_1$ and $y$ are along the transverse directions)
\bea
\bma \xi_1 \\ \xi_3 \ema= \bma x \cos\chi+z \sin\chi \\ -t-x \sin\chi+z \cos\chi \ema
\eea
and let the wave profile be specified by a function $g(\xi_1,y,\xi_3)$. The force-free equations are then 
\begin{widetext}
\begin{align}
-\frac{\partial ^2\psi _1}{\partial t^2}+\frac{\partial ^2\psi _1}{\partial z^2}=&
-\frac{\sin \chi }{2 }\left\{  h_{\times}\left( \sin 2 \chi \left(\frac{\partial^2 g}{\partial \xi^2_3}-\frac{\partial^2 g}{\partial \xi^2_1}\right)-2\cos 2 \chi \frac{\partial^2 g}{\partial \xi_1 \partial \xi_3}\right)
+2h_+  \cos \chi \left( \cos \chi  \frac{\partial^2 g}{\partial y\partial \xi_3}+  \sin \chi \frac{\partial^2 g}{\partial \xi_1 \partial y}\right) \right\}\,,
\label{eq:GenericFFE1} \\
-\frac{\partial ^2\psi _2}{\partial t^2}+\frac{\partial ^2\psi _2}{\partial x^2}+\frac{\partial ^2\psi _2}{\partial y^2}+\frac{\partial ^2\psi _2}{\partial z^2}=&
-\sin\chi \left\{-h_{\times}   \left(\cos\chi \frac{\partial^2 g}{\partial \xi_3 \partial y}+\sin\chi   \frac{\partial^2 g}{\partial \xi_1 \partial y}\right)
+h_+  \left(  \sin \chi\left(\frac{\partial^2 g}{\partial \xi^2_3}+\frac{\partial^2 g}{\partial y^2}\right)- \cos\chi  \frac{\partial^2 g}{\partial \xi_1 \partial \xi_3}\right)\right\}\,. \label{eq:GenericFFE2}
\end{align}
\end{widetext}
Taking $g(\xi_1,y,\xi_3)=\cos (\omega \xi_3+\phi_0)$, these generic expressions reduce to the sinusoidal expressions \eqref{eq:FFE1} and \eqref{eq:FFE2}. 


\begin{thebibliography}{58}
\expandafter\ifx\csname natexlab\endcsname\relax\def\natexlab#1{#1}\fi
\expandafter\ifx\csname bibnamefont\endcsname\relax
  \def\bibnamefont#1{#1}\fi
\expandafter\ifx\csname bibfnamefont\endcsname\relax
  \def\bibfnamefont#1{#1}\fi
\expandafter\ifx\csname citenamefont\endcsname\relax
  \def\citenamefont#1{#1}\fi
\expandafter\ifx\csname url\endcsname\relax
  \def\url#1{\texttt{#1}}\fi
\expandafter\ifx\csname urlprefix\endcsname\relax\def\urlprefix{URL }\fi
\providecommand{\bibinfo}[2]{#2}
\providecommand{\eprint}[2][]{\url{#2}}

\bibitem[{\citenamefont{{Abbott} et~al.}(2016)\citenamefont{{Abbott}, {Abbott},
  {Abbott}, {Abernathy}, {Acernese}, {Ackley}, {Adams}, {Adams}, {Addesso},
  {Adhikari} et~al.}}]{2016PhRvL.116f1102A}
\bibinfo{author}{\bibfnamefont{B.~P.} \bibnamefont{{Abbott}}},
  \bibinfo{author}{\bibfnamefont{R.}~\bibnamefont{{Abbott}}},
  \bibinfo{author}{\bibfnamefont{T.~D.} \bibnamefont{{Abbott}}},
  \bibinfo{author}{\bibfnamefont{M.~R.} \bibnamefont{{Abernathy}}},
  \bibinfo{author}{\bibfnamefont{F.}~\bibnamefont{{Acernese}}},
  \bibinfo{author}{\bibfnamefont{K.}~\bibnamefont{{Ackley}}},
  \bibinfo{author}{\bibfnamefont{C.}~\bibnamefont{{Adams}}},
  \bibinfo{author}{\bibfnamefont{T.}~\bibnamefont{{Adams}}},
  \bibinfo{author}{\bibfnamefont{P.}~\bibnamefont{{Addesso}}},
  \bibinfo{author}{\bibfnamefont{R.~X.} \bibnamefont{{Adhikari}}},
  \bibnamefont{et~al.}, \bibinfo{journal}{Physical Review Letters}
  \textbf{\bibinfo{volume}{116}}, \bibinfo{eid}{061102} (\bibinfo{year}{2016}),
  \eprint{1602.03837}.

\bibitem[{\citenamefont{Grischuk}(1976)}]{Grischuk}
\bibinfo{author}{\bibfnamefont{L.~P.} \bibnamefont{Grischuk}},
  \bibinfo{journal}{Sov. Phys. JETP Lett} \textbf{\bibinfo{volume}{23}},
  \bibinfo{pages}{293} (\bibinfo{year}{1976}).

\bibitem[{\citenamefont{{Maggiore}}(2000)}]{Maggiore00}
\bibinfo{author}{\bibfnamefont{M.}~\bibnamefont{{Maggiore}}},
  \bibinfo{journal}{Phys. Rep.} \textbf{\bibinfo{volume}{331}},
  \bibinfo{pages}{283} (\bibinfo{year}{2000}), \eprint{gr-qc/9909001}.

\bibitem[{\citenamefont{{Giovannini}}(1999)}]{1999PhRvD..60l3511G}
\bibinfo{author}{\bibfnamefont{M.}~\bibnamefont{{Giovannini}}},
  \bibinfo{journal}{\prd} \textbf{\bibinfo{volume}{60}}, \bibinfo{eid}{123511}
  (\bibinfo{year}{1999}), \eprint{astro-ph/9903004}.

\bibitem[{\citenamefont{{Garc{\'{\i}}a-Bellido} and
  {Figueroa}}(2007)}]{2007PhRvL..98f1302G}
\bibinfo{author}{\bibfnamefont{J.}~\bibnamefont{{Garc{\'{\i}}a-Bellido}}}
  \bibnamefont{and} \bibinfo{author}{\bibfnamefont{D.~G.}
  \bibnamefont{{Figueroa}}}, \bibinfo{journal}{Physical Review Letters}
  \textbf{\bibinfo{volume}{98}}, \bibinfo{eid}{061302} (\bibinfo{year}{2007}),
  \eprint{astro-ph/0701014}.

\bibitem[{\citenamefont{{Easther} et~al.}(2007)\citenamefont{{Easther},
  {Giblin}, and {Lim}}}]{Easther07}
\bibinfo{author}{\bibfnamefont{R.}~\bibnamefont{{Easther}}},
  \bibinfo{author}{\bibfnamefont{J.~T.} \bibnamefont{{Giblin}},
  \bibfnamefont{Jr.}}, \bibnamefont{and} \bibinfo{author}{\bibfnamefont{E.~A.}
  \bibnamefont{{Lim}}}, \bibinfo{journal}{Physical Review Letters}
  \textbf{\bibinfo{volume}{99}}, \bibinfo{eid}{221301} (\bibinfo{year}{2007}),
  \eprint{astro-ph/0612294}.

\bibitem[{\citenamefont{{Caprini} et~al.}(2009)\citenamefont{{Caprini},
  {Durrer}, and {Servant}}}]{Caprini09a}
\bibinfo{author}{\bibfnamefont{C.}~\bibnamefont{{Caprini}}},
  \bibinfo{author}{\bibfnamefont{R.}~\bibnamefont{{Durrer}}}, \bibnamefont{and}
  \bibinfo{author}{\bibfnamefont{G.}~\bibnamefont{{Servant}}},
  \bibinfo{journal}{JCAP} \textbf{\bibinfo{volume}{12}}, \bibinfo{eid}{024}
  (\bibinfo{year}{2009}), \eprint{0909.0622}.

\bibitem[{\citenamefont{{Copeland} et~al.}(2009)\citenamefont{{Copeland},
  {Mulryne}, {Nunes}, and {Shaeri}}}]{Copeland09}
\bibinfo{author}{\bibfnamefont{E.~J.} \bibnamefont{{Copeland}}},
  \bibinfo{author}{\bibfnamefont{D.~J.} \bibnamefont{{Mulryne}}},
  \bibinfo{author}{\bibfnamefont{N.~J.} \bibnamefont{{Nunes}}},
  \bibnamefont{and} \bibinfo{author}{\bibfnamefont{M.}~\bibnamefont{{Shaeri}}},
  \bibinfo{journal}{\prd} \textbf{\bibinfo{volume}{79}}, \bibinfo{eid}{023508}
  (\bibinfo{year}{2009}), \eprint{0810.0104}.

\bibitem[{\citenamefont{{Caldwell} et~al.}(1996)\citenamefont{{Caldwell},
  {Battye}, and {Shellard}}}]{Caldwell96}
\bibinfo{author}{\bibfnamefont{R.~R.} \bibnamefont{{Caldwell}}},
  \bibinfo{author}{\bibfnamefont{R.~A.} \bibnamefont{{Battye}}},
  \bibnamefont{and} \bibinfo{author}{\bibfnamefont{E.~P.~S.}
  \bibnamefont{{Shellard}}}, \bibinfo{journal}{\prd}
  \textbf{\bibinfo{volume}{54}}, \bibinfo{pages}{7146} (\bibinfo{year}{1996}),
  \eprint{astro-ph/9607130}.

\bibitem[{\citenamefont{{Leblond} et~al.}(2009)\citenamefont{{Leblond},
  {Shlaer}, and {Siemens}}}]{Leblond09}
\bibinfo{author}{\bibfnamefont{L.}~\bibnamefont{{Leblond}}},
  \bibinfo{author}{\bibfnamefont{B.}~\bibnamefont{{Shlaer}}}, \bibnamefont{and}
  \bibinfo{author}{\bibfnamefont{X.}~\bibnamefont{{Siemens}}},
  \bibinfo{journal}{\prd} \textbf{\bibinfo{volume}{79}}, \bibinfo{eid}{123519}
  (\bibinfo{year}{2009}), \eprint{0903.4686}.

\bibitem[{\citenamefont{{Seahra} et~al.}(2005)\citenamefont{{Seahra},
  {Clarkson}, and {Maartens}}}]{Seahra05}
\bibinfo{author}{\bibfnamefont{S.~S.} \bibnamefont{{Seahra}}},
  \bibinfo{author}{\bibfnamefont{C.}~\bibnamefont{{Clarkson}}},
  \bibnamefont{and}
  \bibinfo{author}{\bibfnamefont{R.}~\bibnamefont{{Maartens}}},
  \bibinfo{journal}{Physical Review Letters} \textbf{\bibinfo{volume}{94}},
  \bibinfo{eid}{121302} (\bibinfo{year}{2005}), \eprint{gr-qc/0408032}.

\bibitem[{\citenamefont{{Clarkson} and {Seahra}}(2007)}]{Clarkson07}
\bibinfo{author}{\bibfnamefont{C.}~\bibnamefont{{Clarkson}}} \bibnamefont{and}
  \bibinfo{author}{\bibfnamefont{S.~S.} \bibnamefont{{Seahra}}},
  \bibinfo{journal}{Classical and Quantum Gravity}
  \textbf{\bibinfo{volume}{24}}, \bibinfo{pages}{F33} (\bibinfo{year}{2007}),
  \eprint{astro-ph/0610470}.

\bibitem[{\citenamefont{{Akutsu} et~al.}(2008)\citenamefont{{Akutsu},
  {Kawamura}, {Nishizawa}, {Arai}, {Yamamoto}, {Tatsumi}, {Nagano}, {Nishida},
  {Chiba}, {Takahashi} et~al.}}]{2008PhRvL.101j1101A}
\bibinfo{author}{\bibfnamefont{T.}~\bibnamefont{{Akutsu}}},
  \bibinfo{author}{\bibfnamefont{S.}~\bibnamefont{{Kawamura}}},
  \bibinfo{author}{\bibfnamefont{A.}~\bibnamefont{{Nishizawa}}},
  \bibinfo{author}{\bibfnamefont{K.}~\bibnamefont{{Arai}}},
  \bibinfo{author}{\bibfnamefont{K.}~\bibnamefont{{Yamamoto}}},
  \bibinfo{author}{\bibfnamefont{D.}~\bibnamefont{{Tatsumi}}},
  \bibinfo{author}{\bibfnamefont{S.}~\bibnamefont{{Nagano}}},
  \bibinfo{author}{\bibfnamefont{E.}~\bibnamefont{{Nishida}}},
  \bibinfo{author}{\bibfnamefont{T.}~\bibnamefont{{Chiba}}},
  \bibinfo{author}{\bibfnamefont{R.}~\bibnamefont{{Takahashi}}},
  \bibnamefont{et~al.}, \bibinfo{journal}{Physical Review Letters}
  \textbf{\bibinfo{volume}{101}}, \bibinfo{eid}{101101} (\bibinfo{year}{2008}),
  \eprint{0803.4094}.

\bibitem[{\citenamefont{{Cruise} and
  {Ingley}}(2006{\natexlab{a}})}]{2006CQGra..23.6185C}
\bibinfo{author}{\bibfnamefont{A.~M.} \bibnamefont{{Cruise}}} \bibnamefont{and}
  \bibinfo{author}{\bibfnamefont{R.~M.~J.} \bibnamefont{{Ingley}}},
  \bibinfo{journal}{Classical and Quantum Gravity}
  \textbf{\bibinfo{volume}{23}}, \bibinfo{pages}{6185}
  (\bibinfo{year}{2006}{\natexlab{a}}).

\bibitem[{\citenamefont{{Yu} et~al.}(In Prep.)\citenamefont{{Yu}, {Zhao},
  {Blair}, {Blair}, {Liu}, {Fang}, {Danilishin}, {Zhang}, and
  {Ju}}}]{TiltInPrep}
\bibinfo{author}{\bibfnamefont{S.}~\bibnamefont{{Yu}}},
  \bibinfo{author}{\bibfnamefont{C.}~\bibnamefont{{Zhao}}},
  \bibinfo{author}{\bibfnamefont{D.}~\bibnamefont{{Blair}}},
  \bibinfo{author}{\bibfnamefont{C.}~\bibnamefont{{Blair}}},
  \bibinfo{author}{\bibfnamefont{J.}~\bibnamefont{{Liu}}},
  \bibinfo{author}{\bibfnamefont{Q.}~\bibnamefont{{Fang}}},
  \bibinfo{author}{\bibfnamefont{S.~L.} \bibnamefont{{Danilishin}}},
  \bibinfo{author}{\bibfnamefont{F.}~\bibnamefont{{Zhang}}}, \bibnamefont{and}
  \bibinfo{author}{\bibfnamefont{L.}~\bibnamefont{{Ju}}} (\bibinfo{year}{In
  Prep.}).

\bibitem[{\citenamefont{{Gertsenshtein}}(1962)}]{Gertsenshtein62}
\bibinfo{author}{\bibfnamefont{M.~E.} \bibnamefont{{Gertsenshtein}}},
  \bibinfo{journal}{JETP} \textbf{\bibinfo{volume}{14}}, \bibinfo{pages}{84}
  (\bibinfo{year}{1962}).

\bibitem[{\citenamefont{{Zipoy}}(1966)}]{1966PhRv..142..825Z}
\bibinfo{author}{\bibfnamefont{D.~M.} \bibnamefont{{Zipoy}}},
  \bibinfo{journal}{Physical Review} \textbf{\bibinfo{volume}{142}},
  \bibinfo{pages}{825} (\bibinfo{year}{1966}).

\bibitem[{\citenamefont{{Braginsky} and {Mensky}}(1972)}]{1972GReGr...3..401B}
\bibinfo{author}{\bibfnamefont{V.~B.} \bibnamefont{{Braginsky}}}
  \bibnamefont{and} \bibinfo{author}{\bibfnamefont{M.~B.}
  \bibnamefont{{Mensky}}}, \bibinfo{journal}{General Relativity and
  Gravitation} \textbf{\bibinfo{volume}{3}}, \bibinfo{pages}{401}
  (\bibinfo{year}{1972}).

\bibitem[{\citenamefont{{Cruise}}(1983)}]{1983MNRAS.204..485C}
\bibinfo{author}{\bibfnamefont{A.~M.} \bibnamefont{{Cruise}}},
  \bibinfo{journal}{Mon. Not. R. Astron. Soc.} \textbf{\bibinfo{volume}{204}},
  \bibinfo{pages}{485} (\bibinfo{year}{1983}).

\bibitem[{\citenamefont{{Cruise}}(2000)}]{2000CQGra..17.2525C}
\bibinfo{author}{\bibfnamefont{A.~M.} \bibnamefont{{Cruise}}},
  \bibinfo{journal}{Classical and Quantum Gravity}
  \textbf{\bibinfo{volume}{17}}, \bibinfo{pages}{2525} (\bibinfo{year}{2000}).

\bibitem[{\citenamefont{Li et~al.}(2009)\citenamefont{Li, Yang, Fang, Baker,
  Stephenson, and Wen}}]{Li:2009zzy}
\bibinfo{author}{\bibfnamefont{F.}~\bibnamefont{Li}},
  \bibinfo{author}{\bibfnamefont{N.}~\bibnamefont{Yang}},
  \bibinfo{author}{\bibfnamefont{Z.}~\bibnamefont{Fang}},
  \bibinfo{author}{\bibfnamefont{R.~M.~L.} \bibnamefont{Baker},
  \bibfnamefont{Jr.}}, \bibinfo{author}{\bibfnamefont{G.~V.}
  \bibnamefont{Stephenson}}, \bibnamefont{and}
  \bibinfo{author}{\bibfnamefont{H.}~\bibnamefont{Wen}},
  \bibinfo{journal}{Phys. Rev.} \textbf{\bibinfo{volume}{D80}},
  \bibinfo{pages}{064013} (\bibinfo{year}{2009}), \eprint{0909.4118}.

\bibitem[{\citenamefont{{Macedo} and {Nelson}}(1983)}]{1983PhRvD..28.2382M}
\bibinfo{author}{\bibfnamefont{P.~G.} \bibnamefont{{Macedo}}} \bibnamefont{and}
  \bibinfo{author}{\bibfnamefont{A.~H.} \bibnamefont{{Nelson}}},
  \bibinfo{journal}{\prd} \textbf{\bibinfo{volume}{28}}, \bibinfo{pages}{2382}
  (\bibinfo{year}{1983}).

\bibitem[{\citenamefont{{Macedo} and {Nelson}}(1990)}]{1990ApJ...362..584M}
\bibinfo{author}{\bibfnamefont{P.}~\bibnamefont{{Macedo}}} \bibnamefont{and}
  \bibinfo{author}{\bibfnamefont{H.}~\bibnamefont{{Nelson}}},
  \bibinfo{journal}{\apj} \textbf{\bibinfo{volume}{362}}, \bibinfo{pages}{584}
  (\bibinfo{year}{1990}).

\bibitem[{\citenamefont{{Greco} and {Seta}}(1998)}]{1998CQGra..15.3655G}
\bibinfo{author}{\bibfnamefont{A.}~\bibnamefont{{Greco}}} \bibnamefont{and}
  \bibinfo{author}{\bibfnamefont{L.}~\bibnamefont{{Seta}}},
  \bibinfo{journal}{Classical and Quantum Gravity}
  \textbf{\bibinfo{volume}{15}}, \bibinfo{pages}{3655} (\bibinfo{year}{1998}).

\bibitem[{\citenamefont{{Brodin} and {Marklund}}(1999)}]{1999PhRvL..82.3012B}
\bibinfo{author}{\bibfnamefont{G.}~\bibnamefont{{Brodin}}} \bibnamefont{and}
  \bibinfo{author}{\bibfnamefont{M.}~\bibnamefont{{Marklund}}},
  \bibinfo{journal}{Physical Review Letters} \textbf{\bibinfo{volume}{82}},
  \bibinfo{pages}{3012} (\bibinfo{year}{1999}), \eprint{astro-ph/9810128}.

\bibitem[{\citenamefont{{Marklund} et~al.}(2000)\citenamefont{{Marklund},
  {Brodin}, and {Dunsby}}}]{2000ApJ...536..875M}
\bibinfo{author}{\bibfnamefont{M.}~\bibnamefont{{Marklund}}},
  \bibinfo{author}{\bibfnamefont{G.}~\bibnamefont{{Brodin}}}, \bibnamefont{and}
  \bibinfo{author}{\bibfnamefont{P.~K.~S.} \bibnamefont{{Dunsby}}},
  \bibinfo{journal}{\apj} \textbf{\bibinfo{volume}{536}}, \bibinfo{pages}{875}
  (\bibinfo{year}{2000}), \eprint{astro-ph/9907350}.

\bibitem[{\citenamefont{{Servin} et~al.}(2000)\citenamefont{{Servin}, {Brodin},
  {Bradley}, and {Marklund}}}]{2000PhRvE..62.8493S}
\bibinfo{author}{\bibfnamefont{M.}~\bibnamefont{{Servin}}},
  \bibinfo{author}{\bibfnamefont{G.}~\bibnamefont{{Brodin}}},
  \bibinfo{author}{\bibfnamefont{M.}~\bibnamefont{{Bradley}}},
  \bibnamefont{and}
  \bibinfo{author}{\bibfnamefont{M.}~\bibnamefont{{Marklund}}},
  \bibinfo{journal}{\pre} \textbf{\bibinfo{volume}{62}}, \bibinfo{pages}{8493}
  (\bibinfo{year}{2000}), \eprint{physics/9910029}.

\bibitem[{\citenamefont{{Papadopoulos}
  et~al.}(2001)\citenamefont{{Papadopoulos}, {Stergioulas}, {Vlahos}, and
  {Kuijpers}}}]{2001A&A...377..701P}
\bibinfo{author}{\bibfnamefont{D.}~\bibnamefont{{Papadopoulos}}},
  \bibinfo{author}{\bibfnamefont{N.}~\bibnamefont{{Stergioulas}}},
  \bibinfo{author}{\bibfnamefont{L.}~\bibnamefont{{Vlahos}}}, \bibnamefont{and}
  \bibinfo{author}{\bibfnamefont{J.}~\bibnamefont{{Kuijpers}}},
  \bibinfo{journal}{Astron. Astrophys.} \textbf{\bibinfo{volume}{377}},
  \bibinfo{pages}{701} (\bibinfo{year}{2001}), \eprint{astro-ph/0107043}.

\bibitem[{\citenamefont{{Servin} and {Brodin}}(2003)}]{2003PhRvD..68d4017S}
\bibinfo{author}{\bibfnamefont{M.}~\bibnamefont{{Servin}}} \bibnamefont{and}
  \bibinfo{author}{\bibfnamefont{G.}~\bibnamefont{{Brodin}}},
  \bibinfo{journal}{\prd} \textbf{\bibinfo{volume}{68}}, \bibinfo{eid}{044017}
  (\bibinfo{year}{2003}), \eprint{gr-qc/0302039}.

\bibitem[{\citenamefont{Duez et~al.}(2005)\citenamefont{Duez, Liu, Shapiro, and
  Stephens}}]{Duez2005}
\bibinfo{author}{\bibfnamefont{M.~D.} \bibnamefont{Duez}},
  \bibinfo{author}{\bibfnamefont{Y.~T.} \bibnamefont{Liu}},
  \bibinfo{author}{\bibfnamefont{S.~L.} \bibnamefont{Shapiro}},
  \bibnamefont{and} \bibinfo{author}{\bibfnamefont{B.~C.}
  \bibnamefont{Stephens}}, \bibinfo{journal}{Physical Review D}
  \textbf{\bibinfo{volume}{72}}, \bibinfo{eid}{024028}
  (pages~\bibinfo{numpages}{21}) (\bibinfo{year}{2005}), ISSN
  \bibinfo{issn}{1550-7998}, \eprint{astro-ph/0503420},
  \urlprefix\url{http://link.aps.org/doi/10.1103/PhysRevD.72.024028}.

\bibitem[{\citenamefont{Goldreich and Julian}(1969)}]{Goldreich:1969sb}
\bibinfo{author}{\bibfnamefont{P.}~\bibnamefont{Goldreich}} \bibnamefont{and}
  \bibinfo{author}{\bibfnamefont{W.~H.} \bibnamefont{Julian}},
  \bibinfo{journal}{Astrophys.J.} \textbf{\bibinfo{volume}{157}},
  \bibinfo{pages}{869} (\bibinfo{year}{1969}).

\bibitem[{\citenamefont{{Blandford} and {Znajek}}(1977)}]{1977MNRAS.179..433B}
\bibinfo{author}{\bibfnamefont{R.~D.} \bibnamefont{{Blandford}}}
  \bibnamefont{and} \bibinfo{author}{\bibfnamefont{R.~L.}
  \bibnamefont{{Znajek}}}, \bibinfo{journal}{Mon. Not. Roy. Astron. Soc.}
  \textbf{\bibinfo{volume}{179}}, \bibinfo{pages}{433} (\bibinfo{year}{1977}).

\bibitem[{\citenamefont{Zhang et~al.}(2015)\citenamefont{Zhang, McWilliams, and
  Pfeiffer}}]{Zhang:2015aga}
\bibinfo{author}{\bibfnamefont{F.}~\bibnamefont{Zhang}},
  \bibinfo{author}{\bibfnamefont{S.~T.} \bibnamefont{McWilliams}},
  \bibnamefont{and} \bibinfo{author}{\bibfnamefont{H.~P.}
  \bibnamefont{Pfeiffer}}, \bibinfo{journal}{Phys. Rev.}
  \textbf{\bibinfo{volume}{D92}}, \bibinfo{pages}{024049}
  (\bibinfo{year}{2015}), \eprint{1501.05394}.

\bibitem[{\citenamefont{Gralla and Jacobson}(2014)}]{Gralla:2014yja}
\bibinfo{author}{\bibfnamefont{S.~E.} \bibnamefont{Gralla}} \bibnamefont{and}
  \bibinfo{author}{\bibfnamefont{T.}~\bibnamefont{Jacobson}},
  \bibinfo{journal}{Mon. Not. Roy. Astron. Soc.}
  \textbf{\bibinfo{volume}{445}}, \bibinfo{pages}{2500} (\bibinfo{year}{2014}),
  \eprint{1401.6159}.

\bibitem[{\citenamefont{{Uchida}}(1997{\natexlab{a}})}]{1997PhRvE..56.2181U}
\bibinfo{author}{\bibfnamefont{T.}~\bibnamefont{{Uchida}}},
  \bibinfo{journal}{\pre} \textbf{\bibinfo{volume}{56}}, \bibinfo{pages}{2181}
  (\bibinfo{year}{1997}{\natexlab{a}}).

\bibitem[{\citenamefont{{Uchida}}(1997{\natexlab{b}})}]{1997PhRvE..56.2198U}
\bibinfo{author}{\bibfnamefont{T.}~\bibnamefont{{Uchida}}},
  \bibinfo{journal}{\pre} \textbf{\bibinfo{volume}{56}}, \bibinfo{pages}{2198}
  (\bibinfo{year}{1997}{\natexlab{b}}).

\bibitem[{\citenamefont{{Yang} and {Zhang}}(2016)}]{2016ApJ...817..183Y}
\bibinfo{author}{\bibfnamefont{H.}~\bibnamefont{{Yang}}} \bibnamefont{and}
  \bibinfo{author}{\bibfnamefont{F.}~\bibnamefont{{Zhang}}},
  \bibinfo{journal}{\apj} \textbf{\bibinfo{volume}{817}}, \bibinfo{eid}{183}
  (\bibinfo{year}{2016}), \eprint{1508.02119}.

\bibitem[{\citenamefont{{Gralla} and {Zimmerman}}(2016)}]{2016arXiv160309693G}
\bibinfo{author}{\bibfnamefont{S.~E.} \bibnamefont{{Gralla}}} \bibnamefont{and}
  \bibinfo{author}{\bibfnamefont{P.}~\bibnamefont{{Zimmerman}}},
  \bibinfo{journal}{ArXiv e-prints}  (\bibinfo{year}{2016}),
  \eprint{1603.09693}.

\bibitem[{\citenamefont{Brennan et~al.}(2013)\citenamefont{Brennan, Gralla, and
  Jacobson}}]{Brennan:2013jla}
\bibinfo{author}{\bibfnamefont{T.~D.} \bibnamefont{Brennan}},
  \bibinfo{author}{\bibfnamefont{S.~E.} \bibnamefont{Gralla}},
  \bibnamefont{and} \bibinfo{author}{\bibfnamefont{T.}~\bibnamefont{Jacobson}},
  \bibinfo{journal}{Class.Quant.Grav.} \textbf{\bibinfo{volume}{30}},
  \bibinfo{pages}{195012} (\bibinfo{year}{2013}), \eprint{1305.6890}.

\bibitem[{\citenamefont{Yang and Zhang}(2014)}]{Yang:2014zva}
\bibinfo{author}{\bibfnamefont{H.}~\bibnamefont{Yang}} \bibnamefont{and}
  \bibinfo{author}{\bibfnamefont{F.}~\bibnamefont{Zhang}},
  \bibinfo{journal}{Phys.Rev.} \textbf{\bibinfo{volume}{D90}},
  \bibinfo{pages}{104022} (\bibinfo{year}{2014}), \eprint{1406.4602}.

\bibitem[{\citenamefont{Schmidt}(1979)}]{PlasmaBook}
\bibinfo{author}{\bibfnamefont{G.}~\bibnamefont{Schmidt}},
  \emph{\bibinfo{title}{Physics of high temperature plasmas}}
  (\bibinfo{publisher}{Academic Press, Inc}, \bibinfo{address}{New York},
  \bibinfo{year}{1979}), ISBN \bibinfo{isbn}{0126266603}.

\bibitem[{\citenamefont{{Lupanov}}(1967)}]{Lupanov67}
\bibinfo{author}{\bibfnamefont{G.~A.} \bibnamefont{{Lupanov}}},
  \bibinfo{journal}{JETP} \textbf{\bibinfo{volume}{25}}, \bibinfo{pages}{76}
  (\bibinfo{year}{1967}).

\bibitem[{\citenamefont{{de Logi} and {Mickelson}}(1977)}]{1977PhRvD..16.2915D}
\bibinfo{author}{\bibfnamefont{W.~K.} \bibnamefont{{de Logi}}}
  \bibnamefont{and} \bibinfo{author}{\bibfnamefont{A.~R.}
  \bibnamefont{{Mickelson}}}, \bibinfo{journal}{\prd}
  \textbf{\bibinfo{volume}{16}}, \bibinfo{pages}{2915} (\bibinfo{year}{1977}).

\bibitem[{\citenamefont{{Li} et~al.}(2013)\citenamefont{{Li}, {Wen}, and
  {Fang}}}]{1674-1056-22-12-120402}
\bibinfo{author}{\bibfnamefont{F.-Y.} \bibnamefont{{Li}}},
  \bibinfo{author}{\bibfnamefont{H.}~\bibnamefont{{Wen}}}, \bibnamefont{and}
  \bibinfo{author}{\bibfnamefont{Z.-Y.} \bibnamefont{{Fang}}},
  \bibinfo{journal}{Chinese Physics B} \textbf{\bibinfo{volume}{22}},
  \bibinfo{pages}{120402} (\bibinfo{year}{2013}),
  \urlprefix\url{http://stacks.iop.org/1674-1056/22/i=12/a=120402}.

\bibitem[{\citenamefont{Cruise}(2012)}]{Cruise:2012zz}
\bibinfo{author}{\bibfnamefont{A.~M.} \bibnamefont{Cruise}},
  \bibinfo{journal}{Class. Quant. Grav.} \textbf{\bibinfo{volume}{29}},
  \bibinfo{pages}{095003} (\bibinfo{year}{2012}).

\bibitem[{\citenamefont{Beznosko et~al.}(2005)\citenamefont{Beznosko, Blazey,
  Dyshkant, and Rykalin}}]{Beznosko:2005sy}
\bibinfo{author}{\bibfnamefont{D.}~\bibnamefont{Beznosko}},
  \bibinfo{author}{\bibfnamefont{G.}~\bibnamefont{Blazey}},
  \bibinfo{author}{\bibfnamefont{A.}~\bibnamefont{Dyshkant}}, \bibnamefont{and}
  \bibinfo{author}{\bibfnamefont{V.}~\bibnamefont{Rykalin}},
  \bibinfo{journal}{Submitted to: NUCL. INSTRUM. METH. A}
  (\bibinfo{year}{2005}).

\bibitem[{\citenamefont{Dyer}(2001)}]{measurementbook}
\bibinfo{author}{\bibfnamefont{S.~A.} \bibnamefont{Dyer}},
  \emph{\bibinfo{title}{Wiley Survey of Instrumentation and Measurement}}
  (\bibinfo{publisher}{John Wiley \& Sons, Inc}, \bibinfo{address}{New York},
  \bibinfo{year}{2001}).

\bibitem[{\citenamefont{{Boccaletti} et~al.}(1970)\citenamefont{{Boccaletti},
  {Sabbata}, {Fortini}, and {Gualdi}}}]{1970NCimB..70..129B}
\bibinfo{author}{\bibfnamefont{D.}~\bibnamefont{{Boccaletti}}},
  \bibinfo{author}{\bibfnamefont{V.}~\bibnamefont{{Sabbata}}},
  \bibinfo{author}{\bibfnamefont{P.}~\bibnamefont{{Fortini}}},
  \bibnamefont{and} \bibinfo{author}{\bibfnamefont{C.}~\bibnamefont{{Gualdi}}},
  \bibinfo{journal}{Nuovo Cimento B Serie} \textbf{\bibinfo{volume}{70}},
  \bibinfo{pages}{129} (\bibinfo{year}{1970}).

\bibitem[{\citenamefont{De~Logi and Mickelson}(1977)}]{DeLogi:1977qe}
\bibinfo{author}{\bibfnamefont{W.~K.} \bibnamefont{De~Logi}} \bibnamefont{and}
  \bibinfo{author}{\bibfnamefont{A.~R.} \bibnamefont{Mickelson}},
  \bibinfo{journal}{Phys. Rev.} \textbf{\bibinfo{volume}{D16}},
  \bibinfo{pages}{2915} (\bibinfo{year}{1977}).

\bibitem[{\citenamefont{{Pegoraro} et~al.}(1978)\citenamefont{{Pegoraro},
  {Radicati}, {Bernard}, and {Picasso}}}]{1978PhLA...68..165P}
\bibinfo{author}{\bibfnamefont{F.}~\bibnamefont{{Pegoraro}}},
  \bibinfo{author}{\bibfnamefont{L.~A.} \bibnamefont{{Radicati}}},
  \bibinfo{author}{\bibfnamefont{P.}~\bibnamefont{{Bernard}}},
  \bibnamefont{and}
  \bibinfo{author}{\bibfnamefont{E.}~\bibnamefont{{Picasso}}},
  \bibinfo{journal}{Physics Letters A} \textbf{\bibinfo{volume}{68}},
  \bibinfo{pages}{165} (\bibinfo{year}{1978}).

\bibitem[{\citenamefont{{Baierlein}}(1976)}]{1976GReGr...7..583B}
\bibinfo{author}{\bibfnamefont{R.}~\bibnamefont{{Baierlein}}},
  \bibinfo{journal}{General Relativity and Gravitation}
  \textbf{\bibinfo{volume}{7}}, \bibinfo{pages}{583} (\bibinfo{year}{1976}).

\bibitem[{\citenamefont{{Tourrenc} and
  {Crossiord}}(1974)}]{1974NCimB..19..105T}
\bibinfo{author}{\bibfnamefont{P.}~\bibnamefont{{Tourrenc}}} \bibnamefont{and}
  \bibinfo{author}{\bibfnamefont{J.-L.} \bibnamefont{{Crossiord}}},
  \bibinfo{journal}{Nuovo Cimento B Serie} \textbf{\bibinfo{volume}{19}},
  \bibinfo{pages}{105} (\bibinfo{year}{1974}).

\bibitem[{\citenamefont{{Tokuoka}}(1975)}]{1975PThPh..54.1309T}
\bibinfo{author}{\bibfnamefont{T.}~\bibnamefont{{Tokuoka}}},
  \bibinfo{journal}{Progress of Theoretical Physics}
  \textbf{\bibinfo{volume}{54}}, \bibinfo{pages}{1309} (\bibinfo{year}{1975}).

\bibitem[{\citenamefont{{Fakir}}(1993)}]{1993ApJ...418..202F}
\bibinfo{author}{\bibfnamefont{R.}~\bibnamefont{{Fakir}}},
  \bibinfo{journal}{\apj} \textbf{\bibinfo{volume}{418}}, \bibinfo{pages}{202}
  (\bibinfo{year}{1993}).

\bibitem[{\citenamefont{{Cooperstock}}(1968)}]{1968AnPhy..47..173C}
\bibinfo{author}{\bibfnamefont{F.~I.} \bibnamefont{{Cooperstock}}},
  \bibinfo{journal}{Annals of Physics} \textbf{\bibinfo{volume}{47}},
  \bibinfo{pages}{173} (\bibinfo{year}{1968}).

\bibitem[{\citenamefont{{Bergmann}}(1971)}]{1971PhRvL..26.1398B}
\bibinfo{author}{\bibfnamefont{P.~G.} \bibnamefont{{Bergmann}}},
  \bibinfo{journal}{Physical Review Letters} \textbf{\bibinfo{volume}{26}},
  \bibinfo{pages}{1398} (\bibinfo{year}{1971}).

\bibitem[{\citenamefont{{Bertotti} and
  {Catenacci}}(1975)}]{1975GReGr...6..329B}
\bibinfo{author}{\bibfnamefont{B.}~\bibnamefont{{Bertotti}}} \bibnamefont{and}
  \bibinfo{author}{\bibfnamefont{R.}~\bibnamefont{{Catenacci}}},
  \bibinfo{journal}{General Relativity and Gravitation}
  \textbf{\bibinfo{volume}{6}}, \bibinfo{pages}{329} (\bibinfo{year}{1975}).

\bibitem[{\citenamefont{{Cruise} and {Ingley}}(2006{\natexlab{b}})}]{Cruise06}
\bibinfo{author}{\bibfnamefont{A.~M.} \bibnamefont{{Cruise}}} \bibnamefont{and}
  \bibinfo{author}{\bibfnamefont{R.~M.~J.} \bibnamefont{{Ingley}}},
  \bibinfo{journal}{Classical and Quantum Gravity}
  \textbf{\bibinfo{volume}{23}}, \bibinfo{pages}{6185}
  (\bibinfo{year}{2006}{\natexlab{b}}).

\end{thebibliography}

\end{document}